\documentclass[aps,prb,twocolumn,groupedaddress,showpacs]{revtex4}
\usepackage{epsfig}

\newcommand{\summ}{\mathop{\sum}}

\begin{document}

\title{Monte Carlo Simulations of an Extended
Feynman-Kikuchi Model}
\author{A. Zujev and R.T. Scalettar}
\affiliation{Physics Department, University of California, 
Davis, California 95616, USA}

\begin{abstract}
We present Quantum Monte Carlo simulations of a generalization
of the Feynman-Kikuchi model which includes the possibility of
vacancies and interactions between the particles undergoing exchange.
By measuring the winding number (superfluid density) and density
structure factor,
we determine the phase diagram, and show that it exhibits
regions which possess both superfluid and charge ordering.
\end{abstract}

\pacs{
05.30.Jp, % Boson systems
03.75.Hh, % Static properties of condensates; thermodynamical,
%           % statistical and structural properties
% 67.40.Kh, % Boson degeneracy, superfluidity of He4: Thermodyn. props.
% 71.10.Fd, % Lattice fermion models (Hubbard model, etc.)
% 71.30.+h, % Metal-insulator transitions and other electronic transitions
% 02.70.Uu  % Applications of Monte Carlo methods
}
\maketitle

\section{Introduction}

The study of continuum superfluid phase transitions using Quantum Monte Carlo
(QMC) methods has a history which includes path integral simulations
of Helium using realistic interatomic potentials, which capture
$T_\lambda$ in good quantitative agreement with experiment
\cite{ceperley82}, to recent numerical work \cite{newcontinuumsupersims}
focusing on experiments \cite{chan05,moreexpts} which observe
`supersolid' order \cite{supersolidreviews}, the simultaneous presence
of both superfluidity and long range density correlations.

At the same time, related path integral studies of lattice models
(the `boson-Hubbard' Hamiltonian) \cite{fisher89,jaksch98} have been undertaken.
These too have been partially motivated by the issue of supersolid
order,
%% \cite{oldlatticesupersims}, 
but have also been driven by the possibility
of studying the universal conductivity in granular
superconductors \cite{univcond}, 
and superfluid-Mott insulator transitions 
of relevance to optically trapped
atoms \cite{jaksch98}
%% \cite{batrouni90,recentbhsims,opticallatticeexpts}.

Many of these simulations emphasize Feynman's picture of the connection
between the superfluid transition and the increasing entanglement (and
ultimate development of macroscopic `winding' across the whole sample)
of quantum paths as the temperature is lowered.  Indeed, the
superfluid density $\rho_s$ is proportional to the mean square winding
of paths around the lattice\cite{ceperley82}.

However, even before the advent of these large scale QMC
simulations which allow the study of the superfluid transition
exactly, Feynman\cite{feynman53} and
Kikuchi\cite{kikuchi54,kikuchi60} suggested, and studied analytically,
an approximate `classical'
model whose configurations are permutation loops of sites on a $d=3$
lattice.  The partition function they suggested is,
\begin{eqnarray}
Z = \int \prod_{i=1}^N d{\bf r}_i \sum_{\cal P}
\rho({\bf r}_1, {\bf r}_2,\cdots , {\bf r}_N)
\nonumber
\\
\times \,\,
{\rm exp} \, \big[ \,\, - {m_{\rm eff} k_{\rm B} T \over 2 \hbar^2} 
\sum_i ( {\bf r}_i - {\cal P} {\bf r}_i )^2 \,\,\big] \,\,.
%%% \nonumber
\end{eqnarray}
Here $m_{\rm eff}$ is the effective mass of He atoms, and the function
$\rho({\bf r}_1, {\bf r}_2,\cdots , {\bf r}_N)$ is assumed to be
nonvanishing only when the coordinates ${\bf r}_i$ are located on the
sites of a regular lattice (a cubic lattice in the original treatments).
${\cal P}$ refers to a permutation of the coordinates.  A transition to
a `superfluid' phase where a macroscopic number of sites participate in
a single large loop, as the temperature $T$ is lowered, was discovered
and investigated.

Various approximations were employed to determine the properties of this
model, predominantly diagrammatic (series) expansions in the exchange
loops.  To faclitate these analytic treatments, in many of the early
studies the allowed permutations were restricted to ``near-neighbor"
exchange in which the maximum
``distance traveled" by each particle $i$, is only one lattice constant
$d$.  That is,
% $ ({\bf r}_i - {\cal P} {\bf r}_i )^2 \leq  d^2$.
$ |{\bf r}_i - {\cal P} {\bf r}_i | =  d$ for all particles $i$.  
One of the early issues
concerned whether the superfluid transition was third order, as
originally found by Feynman \cite{feynman53}, or second order
\cite{chester54,rice54,matsubara54,terhaar54} and whether the order was
affected by the restriction to near-neighbor exchange.  Another issue
was the behavior of the specific heat, both how to eliminate various
artificial structures (and even negative values) near the phase
transition, and also how to recover the experimentally observed $T^3$
behavior in superfluids 
at low temperatures.  Here the removal of the restriction to
local permutations was found to be crucial \cite{kikuchi60}.

The Feynman-Kikuchi (FK) model is closely related to the duality-transformed
$XY$ model \cite{dasgupta81,elser84}, where the partition function can
also be expressed in terms of sums of closed paths on a lattice.  The
allowed configurations are somewhat different, since in the $XY$ case
path overlap is allowed whereas in the FK model, each 
${\bf r}_i$ appears only once in ${\cal P}{\bf r}_i$.  
However, the energy for paths grows
quadratically with the overlap, so that in practice large overlaps are
unlikely, enhancing the similarities between the partition functions.
This connection is perhaps not so surprising since both models offer
ways to understand the superfluid phase transition.

In this paper we will study the FK model in $d=2$ using
Monte Carlo simulations \cite{elser84}.  Motivated by recent work on
supersolids, we will then suggest a generalization which contains
`vacancies' and interactions between the occupied sites.  We will
determine the nature of the superfluid phase transition, and how it
depends on particle density, and also study the possibility of charge
ordered states arising from the interactions.  The results allow up to
construct the phase diagram of our generalized FK model.

\section{Model and Computational Methods}

We begin by briefly reviewing the motivation for the FK model
which will expose the connection with exact path integral
expressions for the partition function.  Consider the quantum Hamiltonian
for a system of $N$ interacting bosons,
\begin{eqnarray}
\hat H = \sum_{i=1}^N {\hat {\bf p}_i^2 \over 2 m}
+ V(\hat {\bf r}_1, \hat {\bf r}_2, \cdots \hat {\bf r}_N)
\,\, .
%%% \nonumber
\end{eqnarray}
Here $\hat {\bf p}_i$ and
$\hat {\bf r}_i$ are the momentum and position operators.
The partition function is given by,
\begin{eqnarray}
Z &=& {\rm Tr} \, e^{-\hat H / T}
= {\rm Tr} \, \big[ \, e^{-\epsilon \hat H / T}
%%%%%%%  \, e^{-\epsilon \hat H / T}
\cdots
\, e^{-\epsilon \hat H / T} \, \big]
\nonumber
\\
&\approx &
{\rm Tr} \, \big[ e^{-\epsilon \hat K / T}
\, e^{-\epsilon \hat V / T}
%%%%%% \, e^{-\epsilon \hat K / T}
%%%%%% \, e^{-\epsilon \hat V / T}
\cdots
\, e^{-\epsilon \hat K / T}
\, e^{-\epsilon \hat V / T} \, \big]
%%% \nonumber
\end{eqnarray}
where, following the usual the path integral approach 
\cite{feynman65,creutz81}, a small parameter
$\epsilon$ has introduced, the exponential of the full Hamiltonian has
been broken into $M$ pieces with $M\epsilon=1$, and then approximated by
the product of the exponentials of the kinetic and potential energies
individually.  This `Trotter' approximation\cite{trotter,suzuki,fye}
becomes exact in the limit $\epsilon \rightarrow 0$ ($M \rightarrow
\infty$).  We have set Boltzmann's constant $k_{\rm B}=1$ for simplicity.

The trace is evaluated by summing over a complete set of position
eigenstates,  and also inserting additional
complete sets of position eigenstates
throughout the string of incremental imaginary time 
evolution operators.  The potential energy exponentials act
on the eigenstates to give numbers, and the remaining
matrix elements of the kinetic energy
operators are readily computed, yielding,
\begin{eqnarray}
Z &=& \sum_{\cal P}
\, \int \prod_{i=1}^N \prod_{m=1}^M d{\bf r}_i^m  \, e^{-S}
\nonumber
\\
S &=& 
{\epsilon \over T}  
\sum_{m=1}^M V({\bf r}_{1}^m, {\bf r}_2^m, \cdots {\bf r}_{N}^m)
\nonumber
\\
&+& {\epsilon \over T} \sum_{m=1}^M \sum_{i=1}^N
\big[ { {\bf r}_i^{m+1}  - {\bf r}_i^{m} \over \epsilon/T }\big]^2
\,\, .
%% \nonumber
\end{eqnarray}
Here the superscript $m$ is an `imaginary time' index which labels the
point of insertion of the different complete sets of states.  The final
set of positions $\{ {\bf r}_i^M \}$ is constrained to be a permutation
${\cal P}$ of the original positions $\{ {\bf r}_i^1 \}$, as a
consequence of the trace in the definition of the quantum partition
function.  The sum over permutations ${\cal P}$ incorporates the
indistiguishability of the bosonic particles.  This completes the
representation of the partition function as an integral over classical
paths in space and imaginary time.

Examination of this {\it exact} expression for the partition
function, now readily motivates the origin of the FK model.
The function $\rho({\bf r}_1, {\bf r}_2,\cdots , {\bf r}_N)$,
and its restriction to a cubic lattice,
can be thought of as arising from the potential energy terms which
act to tend to localize the particle positions in a regular array.
The {\it single} exponential in the particle positions in the
FK model can be regarded as a truncation of the complete set of
$M$ exponentials for all imaginary times.  Notice that the combination
of the leading factor $\epsilon / T$ and the two such factors in
the denominator of the `kinetic energy' lead to the appearance of the
temperature in the {\it numerator}.  From the viewpoint of the
original path integral, this reflects the fact that as the temperature is
lowered, the paths have more (imaginary) time in which to
propagate, and it becomes increasingly easy for them to permute.
Physically, this then leads to a superfluid phase transition as $T$ is 
lowered. 

In this paper, we will work with the FK model 
on a two dimensional square lattice.  We will denote by $\beta$
the inverse of the prefactor of the sum of the distances traveled by the
individual particles.  That is, our partition function will be,
\begin{eqnarray}
Z = \sum_{{\cal P}} {\rm exp} \big[-{E \over \beta}\big]
 = \sum_{{\cal P}}
{\rm exp} \, \big[- {1 \over \beta}
\sum_i ( {\bf r}_i - {\cal P} {\bf r}_i )^2 \big] 
%% \nonumber
\end{eqnarray}
To be explicit, to a labeling of the sites of the two
dimensional square lattice, illustrated in Fig.~1 (left) we associate 
a permutation ${\cal P}$.  An example is shown in Fig.~1 (right).
This configuration 
contains three nontrivial permutation loops:  The particles at sites 
8 and 9 are in a `two particle' loop, as are the particles
at sites 5 and 18.  Note that in the former each particle moves
a single lattice site, and contributes `1' to the energy, while
the square of the distances traveled by each particle in the latter 
case is 5.  The fourth and fifth rows contain a loop
of five particles (moving distances 
$( {\bf r}_i - {\cal P} {\bf r}_i )^2 =5,1,2,1,$ and $1$)
which extends (`winds') all the way across the lattice in the $x$ direction.
The remaining sites, which share the same labels in the left and right,
correspond physically to particles which have not undergone an
exchange.  The total energy $E$ of the three loops in this
configuration ${\cal P} {\bf r}_i $ is $E=(1+1)+(5+5)+(5+1+2+1+1)=22$.

The periodic boundary conditions produce an ambiguity in the definition
of the energy, since there is more than one way in which to
compute the distance traveled by each 
particle.  We define the energy by choosing the smallest of such
distances.

In order to monitor the superfluid phase transition we measure the
winding across the lattice in the $x$ and $y$ directions.  $W_x$ 
counts the difference between the number of particles which move to the
right across the vertical edge of the lattice and those which move to
the left.  An analogous definition applies for $W_y$ across the top,
horizontal edge.  We define $W^2 = W_x^2+W_y^2$.
In the right panel of Fig.~1, $W_x=1$ and $W_y=0$.

%% REMOVE LOCAL MOVE RESULTS
%% As mentioned in the introduction, in order to facilitate the
%% approximate analytic treatment, early work on the FK model
%% restricted the permutation sum to one in which there are only
%% local exchanges.  
%% The exchange loop involving particles 5 and 18 of Fig.~1 (right) would not
%% be allowed in this case, nor would the five particle loop.
%% After Feynman's original study,
%% it was realized that this restriction
%% leads to certain unphysical results. 
%% Therefore while we will present a few
%% simulations of the restricted model, we will mostly concentrate
%% on the case when all permutations are allowed.
%% REMOVE LOCAL MOVE RESULTS

\begin{figure}[t]
%\epsfig{figure=lattice1.eps,angle=  0,width=3.5cm}
%\hskip1cm
% \epsfig{figure=lattice2.eps,angle=  0,width=3.5cm}
%\epsfig{figure=lattice3.eps,angle=  0,width=3.85cm}
\epsfig{figure=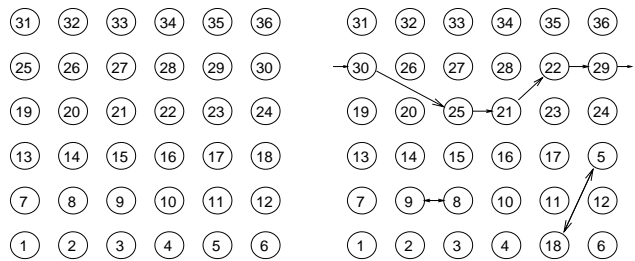,angle=  0,width=9cm}
\caption{Left:  Unpermuted labeling of the sites in a $6 \times 6$ lattice.
Right: Representative configuration in our simulation.  
The energy $E=22$ and the windings $W_x=1, W_y=0$.
See text for explanation.}
\end{figure}

We now introduce our extension of the FK model to allow for
vacancies.  In so doing, we are motivated by recent work on `supersolid'
phases \cite{chan05,newcontinuumsupersims} which are, in fact,
continuations of an extensive history.
%% \cite{oldlatticesupersims,supersolidreviews}.
In the configuration in the right panel of Fig.~1, each site is labeled
by one of the 36 sites in the left panel.  All sites are occupied.
We can define a set of
`vacancies' by assigning `0' to a subset of the sites in the lattice
(Fig. 2, left).  An allowed configuration ${\cal P}({\bf r}_i)$ of the   
system has zeroes at the same sites as the vacant
sites in ${\bf r}_i$ and 
the occupied site labels are permuted.
An example is given in Fig.~2 (right).
By a similar calculation to that described for Fig.~1,
the configuration shown has energy $E = 20$.  %%xx.
Note that we will consider ``annealed'' vacancies:  the 
vacancy density is fixed, but the system is allowed to sample
all possible density locations satisfying that global constraint.

\begin{figure}[t]
%\epsfig{figure=lattice1.eps,angle=  0,width=3.5cm}
%\hskip1cm
%\epsfig{figure=lattice1.eps,angle=  0,width=3.5cm}
\epsfig{figure=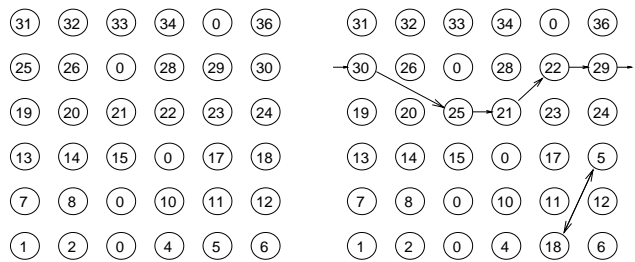,angle=  0,width=9cm}
\caption{Left:  Unpermuted labeling of the sites in a $6 \times 6$ lattice
when the number of bosons $N_b= 31 < 36$.  Empty sites are
labeled by zeroes.
Right: Representative configuration in our simulation of our
extension of the FK model.  
See text for explanation.}
\end{figure}

Once vacancies are allowed, it is possible to introduce additional
terms in the summand of the partition function which control 
their relative positions on the lattice.  We choose to add 
the simplest possible term,
\begin{eqnarray}
E_v &=& +V \sum_{\langle ij\rangle} (\, 1- \delta_{0,{\cal P}{\bf r}_i} \,)
\, (\, 1- \delta_{0,{\cal P}{\bf r}_j} \,)
\nonumber
\\
 &=& +V \sum_{\langle ij\rangle} (\, 1- \delta_{0,{\bf r}_i} \,)
\, (\, 1- \delta_{0,{\bf r}_j} \,)  \,\,\, .
%% \nonumber
\end{eqnarray}
The sum is over neighboring sites $\langle ij\rangle$ of the lattice.
$E_v$ adds $V$ to the energy for each link which connects
sites both of which are occupied.  As emphasized in
Eq.~6, this number is the same whether
the permuted or unpermuted sites are used in the summand, since the
collection of occupied sites before and after the exchanges is the same.

The resulting partition function combines the original FK
exchange term and the new interaction term,
\begin{eqnarray}
Z &=& \sum_{{\cal P}}
{\rm exp} \, \big[ \,\, - {1 \over \beta}
\sum_i ( {\bf r}_i - {\cal P} {\bf r}_i )^2 \,\,
\nonumber
\\
&-& \beta \, V \sum_{\langle ij\rangle} (\, 1- \delta_{0,{\cal P}{\bf r}_i} \,)
\, (\, 1- \delta_{0,{\cal P}{\bf r}_j} \,)
\big] \,\,.
%% \nonumber
\end{eqnarray}
The inverse temperature $\beta$ appears in the interaction term in
its usual place.

We conclude this section by briefly discussing our simulation algorithm.
Our approach is a straightforward implementation of the Metropolis Monte
Carlo method \cite{binder80}.  
We suggest a change in our permutation which consists of
interchanging two, randomly selected, entries ${\cal P}{\bf r}_i$ and
${\cal P}{\bf r}_j$ in the permutation ${\cal P}$.  If a vacancy is moved, 
the list of occupied sites and their permutation
must be changed accordingly.  The resulting change
$\Delta$ in the argument of the exponential appearing in the partition
function is evaluated, and the change is accepted with probability
$p={\rm min}\, (\,1,e^{-\Delta } \,)$.

As with most path integral simulations, such local moves have difficulty
evolving the configuration through phase space at large $\beta$ where the
important paths are dominated by large loops.  We therefore also
introduce `global' moves which shift the elements ${\cal P}{\bf r}_i$ by
one lattice constant for sites $i$ across an entire column or row of the
lattice.  Such moves change the vertical or horizontal winding of the
lattice by $\Delta W = \pm 1$.  In some simulations
such moves have low acceptance rates as the system size increases.  Indeed,
this is a primary limitation of simulations of real 
Helium \cite{ceperley82}.  However, we do not encounter this
difficulty here.

\section{Simulation Results:  Feynman-Kikuchi Model}

We begin by studying the original FK model.  In Fig.~3 we
show data for the mean square winding $\langle W^2 \rangle$ as a
function of $\beta$ for different lattice sizes.  We see that the
winding becomes nonzero as $\beta$ increases, and that the onset of
nonzero winding becomes increasingly sharp as the lattice size grows.
This raw data is suggestive of a critical $\beta_c \approx 0.6$ for the
development of macroscopic loops.
In simulations of the FK model in which only
local exchanges are allowed, on a $d=3$ cubic lattice \cite{elser84}, 
Elser finds $\beta_c \approx 0.69$.  Presumably the higher dimensionality
lowers $\beta_c$ relative to our $d=2$ square lattice while the
restriction to local exchage would tend to raise $\beta_c$.
Hence a rough match of the critical points is plausible.

\begin{figure}[t]
\centerline{\epsfig{figure=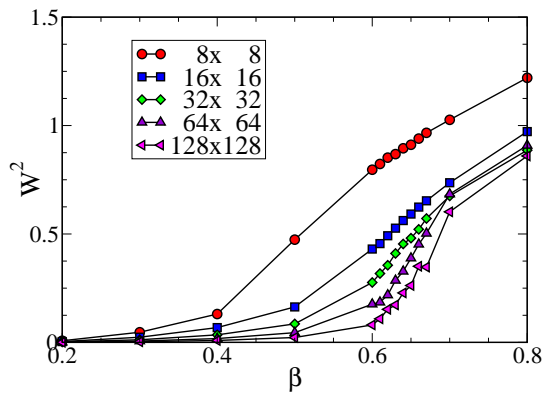,angle=-90,width=8cm}}
\caption{\label{XHO}
Raw data for the mean square winding as a function of
$\beta$ on lattice sizes ranging from $8 \times 8$ to $128 \times 128$.}
\end{figure}

We can make the case for a phase transition, and
determine $\beta_c$ with higher precision, by scaling
our raw data.  We adopt the usual ansatz \cite{cardy88} which postulates
that the dependence of the order parameter on the parameter
controlling the transition and on the lattice size takes the scaling form,
\begin{eqnarray}
W^2 \, (L,\beta) = L^a f\big[ \, L^b (\beta-\beta_c) \, \big]
\,\,.
%% \nonumber
\end{eqnarray}
Here $f$ is a universal (lattice size independent) 
function of its argument, and  $a$ and $b$ are critical exponents.

This scaling form is usefully rewritten as,
\begin{eqnarray}
L^{-a} W^2 \, (L,\beta) = f\big[ \, L^b (\beta-\beta_c) \, \big]
\,\,.
%% \nonumber
\end{eqnarray}
From this expression it is clear that if we scale the order parameter
$\langle W^2 \rangle$ by the lattice size to an appropriate exponent,
$L^{-a} \langle W^2 \rangle$, and plot as a function of the control
parameter $\beta$, all the curves will cross at the universal value
$f(0)$ when $\beta=\beta_c$, regardless of the value of the second
exponent $b$.

Fig.~4 presents the results of the analysis in which the vertical (order
parameter, $\langle W^2 \rangle$) axis alone is scaled.  We observe a
universal crossing of the five curves, and infer $\beta_c = 0.62 \pm
0.01$ and $a=0.61 \pm 0.03$.

\begin{figure}[t]
\centerline{\epsfig{figure=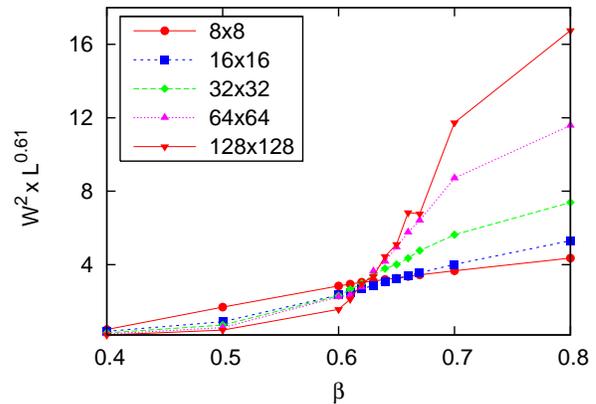,angle=  0,width=8cm}}
\caption{\label{XHOS}
Raw data of Fig.~3 for the mean square winding scaled by
the lattice size.  The intersection of the curves determines 
$\beta_c = 0.62 \pm 0.01$.}
\end{figure}

\begin{figure}[h]
 \includegraphics[width=240pt,angle=0]{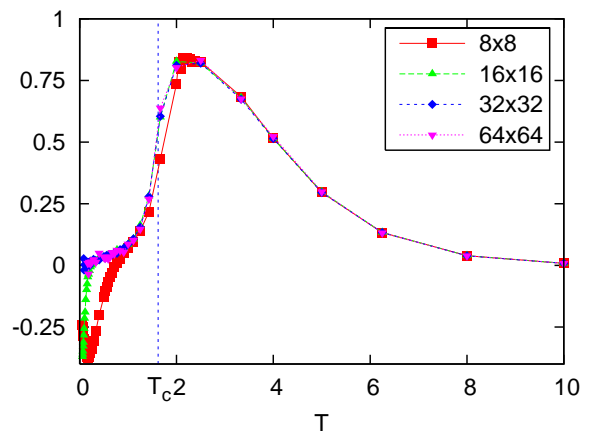}
\caption{\label{CX}Plot of specific heat for the original FK model,
that is, a fully filled lattice with no vacancies and
no interactions.  The value of the critical temperature inferred
from the scaling of the winding (Fig.~4) is indicated as
a vertical dotted line.  The negative values of $C(T)$ at low
$T$ are a finite size artifact, as explained in the text.
}
\end{figure}

We calculate the average energy and specific heat from the
partition function using the standard thermodynamics formulae,
\begin{eqnarray}
  <E> = -\frac{\partial \ln Z}{\partial \beta} 
\hskip0.5in
  C=\frac{\partial E}{\partial T} 
\end{eqnarray}
Fig.~\ref{CX} shows a plot of specific heat as a function of temperature.
We used the temperature rather than $\beta$ as our horizontal
axis, as is more conventional.
We see that there is a peak 
in the specific heat at roughly the same position as the crossing
of the winding $T_c = 1 / \beta_c = 1/ 0.62 = 1.61$.

 The specific heat exhibits a sharp drop, and goes negative, 
close to $T=0$.  
 This is a non-physical effect which appears due to the finite size of the
lattice, as can be seen as follows.
 Starting with the partition function of Eq.~5 we obtain,
\begin{eqnarray}
 E = -\frac{\partial \ln Z}{\partial \beta} 
 = 
 -\frac{1}{\beta^2} \frac{\summ_{n}K e^{- \frac{K}{\beta}}}{Z_0} 
 = 
 -\frac{1}{\beta^2} \; \langle E \rangle 
\end{eqnarray}
 where $Z_0= \summ_{n} 
e^{- \frac{K}{\beta}}$ is the partition function used in Monte Carlo simulations;
 $\langle E \rangle$ is our MC calculation of energy. 
The specific heat
\begin{eqnarray}
 C &=& \frac{\partial E}{\partial T} =\frac{\partial E}{\partial (1/\beta)} 
\nonumber \\
 &=& 
 -\frac{1}{\beta} \langle E \rangle+
\frac{1}{\beta^2} \left( \langle E^2 \rangle 
- {\langle E \rangle}^2 \right) 
\nonumber \\
 &=& -\frac{1}{\beta} B +\frac{1}{\beta^2} A 
\end{eqnarray}
For a finite lattice,
 as $\beta \rightarrow \infty$, $A$ and $B$ approach constant values
given by all permutations 
on the $L \times L$ lattice having equal probability.
As a consequence,
as $\beta \rightarrow \infty$,
 \[ -\frac{1}{\beta} B +\frac{1}{\beta^2} A = \frac{1}{\beta^2} (A- B\beta) \; \rightarrow 0^- \]
 So at large $\beta$, or small $T$, 
the specific heat
 becomes negative, achieves its minimum, 
and goes back to zero, as $T \rightarrow 0$.
%% The reason this artifact is less apparent as the lattice
%% size increases is that **insert explanation if we have one**.
As the lattice size increases, the area of negative specific heat
moves closer to $T=0$. This is because the situation when all permutations have approximately equal probability
occurs when $K/\beta \rightarrow 0$, or  $L^2/\beta \rightarrow 0$,
so $\beta \sim L^2$, or $T \sim L^{-2}$ for this area of negative specific heat.

Despite the crudity of the model, we can, following
Feynman and Kikuchi\cite{feynman53,kikuchi54},  use these results
to infer a rough critical temperature for Helium,
The scaling of the winding gives $\beta_c=0.62$, or $T_c=1.62$.
 To recover physical values for the temperature we note our
unit is
 $T_1 = \hbar^2/k_B m d^2$,
 where $m$ is the mass of boson, and $d$ is lattice spacing.
 For $^4$He, using density of 146 $kg/m^3$, and $d= 3.57 \cdot 10^{-10}m$,
we obtain $T_1=0.95K$.
 This puts our transition temperature around 1.5 K, which is in the same
ballpark as $T_\lambda$ for Helium.
 \newline
  
\begin{figure}[h]
 \includegraphics[width=240pt,angle=0]{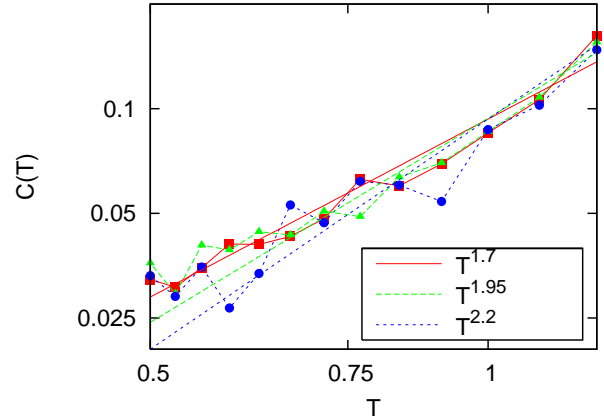}
\caption{\label{CXXLOG} A log-log plot of
specific heat versus temperature yields data consistent with
a straight line of slope $1.95 \pm 0.1$, in agreement with
the expected value, 2.
We used a temperature range $0.5\le T \le 1.25$, where $T$ is small, 
but out of range of the (non-physical) dip in $C$,
}
\end{figure}

We close this section by examining the low $T$ behavior of the specific
heat, since obtaining the proper exponent was the focus of much of
the original work on the FK model.
In three dimensions, where the initial analytic studies were performed,
a linearly dispersing (phonon) mode $E(k) = c k$
gives $C(T) \; \sim \; T^3$ at low $T$.
Here we are working in $d=2$, where instead $C(T) \; \sim \; T^2$.
 Fig.~\ref{CXXLOG} shows an attempt to fit $C(T)$ to a power law.
 The least squares fit 
 gives 
 $C(T) \; \sim \; T^{1.95 \pm 0.1}$
 for lattices of different sizes, 
 which is in reasonable agreement with the prediction based
on a linearly dispersing mode.
(We restrict our fit to temperatures higher than those at which the
finite size artifact negative $C(T)$ values onset.)

%% I REMOVED THE SECTION ON LOCAL CONSTRAINTS.  THE TEXT 
%% IS PRESERVED AT THE BOTTOM OF THE FILE BELOW \end{document}

\section {Simulation Results:  Extended Feynman Kikuchi Model}

We now turn to our generalization of the FK model in
which we allow a lattice with partial filling 
and a repulsive interaction $V$ between nearest neighbor sites.
We study first the special case of half-filling, where it is possible
to have perfect ordering of the vacancies/particles in a 
``checkerboard" pattern.

\subsection {Half-filling: $\langle \rho \rangle \;= \; \frac{1}{2}$}

How does
the introduction of vacancies affect 
$\beta_c^{\,{\rm sf}}$ in the absence of interactions, $V=0$?
(Henceforth in this manuscript we will append 
a superscipt `sf' to $\beta_c$ for the superfluid transition,
to distinguish it from the $\beta_c^{\,{\rm cdw}}$ for charge 
ordering.  See below.)
Figs.~\ref{XH} and \ref{XHS} are the analogues of
Figs.~\ref{XHO} and \ref{XHOS}, and show the unscaled
and scaled winding as a function of $\beta$.
As before, we have done simulations for lattices of different sizes
to perform the finite size scaling.  The crossing occurs at
$\beta_c^{\,{\rm sf}} = 1.50 \pm 0.02$.
By repeating this sweep of $\beta$ for different $V$,
we can compute the superfluid phase boundary $\beta_c^{\,{\rm sf}}(V)$ in the 
$V-\beta$ plane at half-filling.  This is shown in Fig.~\ref{Phases}.

\begin{figure}[t]
\centerline{\epsfig{figure=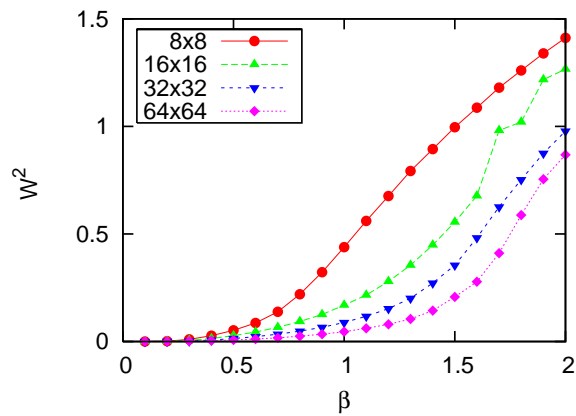,angle=  0,width=8cm}}
\caption{\label{XH} Raw data for 
$\langle W^2 \rangle$ for the extended FK model with half-filling and 
$V=0$.}
\end{figure}

\begin{figure}[t]
\centerline{\epsfig{figure=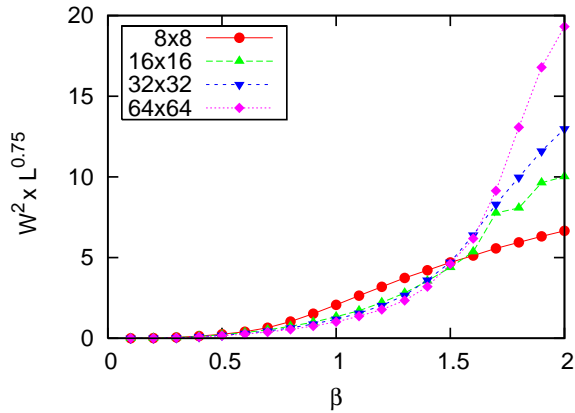,angle=  0,width=8cm}}
\caption{\label{XHS}
Data of Fig. \ref{XH}, scaled.  
We infer $\beta_c^{\,{\rm sf}} = 1.50 \pm 0.02$.
}
\end{figure}

%% SPECIFIC HEAT DISCUSSION REMOVED. THE TEXT 
%% IS PRESERVED AT THE BOTTOM OF THE FILE BELOW \end{document}

A natural question to ask is whether $E_v$ 
induces vacancy/density ordering.
We define the real space density correlation function,
\begin{eqnarray}
c({\bf r}, {\bf r}\,') \; = \; \langle \,\, \left( \rho({\bf r}) - 
\left< \rho \right> \right) \left( \rho({\bf r}\,') - 
\left< \rho \right> \right) \,\, \rangle  \,\,\, ,
\end{eqnarray}
where the density $\rho({\bf r}_i)=1$ if the site $i$ is
occupied and is zero otherwise.
The structure function,
\begin{eqnarray}
  S({\bf q}) = \; \frac{1}{N^2}
\mathop{\sum}_{{\bf r}, {\bf r}\,'}
e^{i {\bf q} \cdot({\bf r} -{\bf r}\,')} c({\bf r}, {\bf r}\,')  \,\,\, ,
\end{eqnarray}
is the Fourier transform of the density correlations.
Here $N$ is number of sites.
At half-filling, the ordering vector ${\bf q}=(\pi,\pi)$.
In a disordered phase where 
$c({\bf r}, {\bf r}\,')$
decays exponentially with $|{\bf r}-{\bf r}'|$,
$S({\bf q})$ will vanish in the thermodynamic limit
(proportional to $1/N$). 
In an ordered phase, $S({\bf q)}$ will go to a constant as the
system size increases, for the appropriate ordering
wave vector ${\bf q}$.
   
\begin{figure}[t]
\centerline{\epsfig{figure=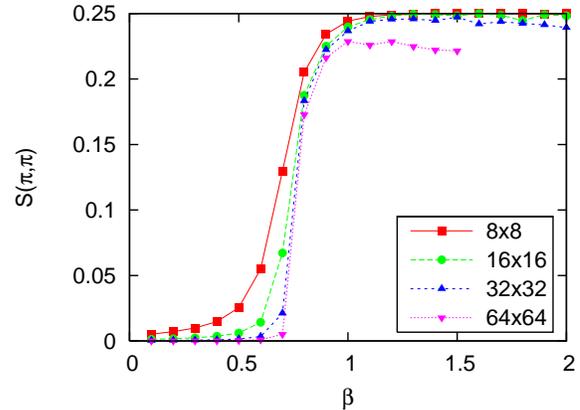,angle=  0,width=8cm}}
\caption{\label{DD}
The structure factor 
$S(\pi,\pi)$ is shown as a function of $\beta$ at half-filling and 
$V=2.5$.}
\end{figure}

\begin{figure}[t]
\centerline{\epsfig{figure=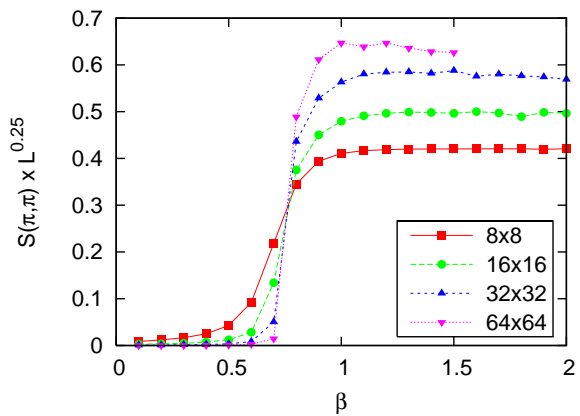,angle=  0,width=8cm}}
\caption{\label{DS}
Scaled data of Fig.~\ref{DD}.  The
crossing point defines $\beta_c^{\,{\rm cdw}}$ for CDW ordering.
The scaling exponent multiplying $S$ on the vertical axis is that
of the $d=2$ Ising model, as discussed in the text.
}
\end{figure}

\begin{figure}[t]
\centerline{\epsfig{figure=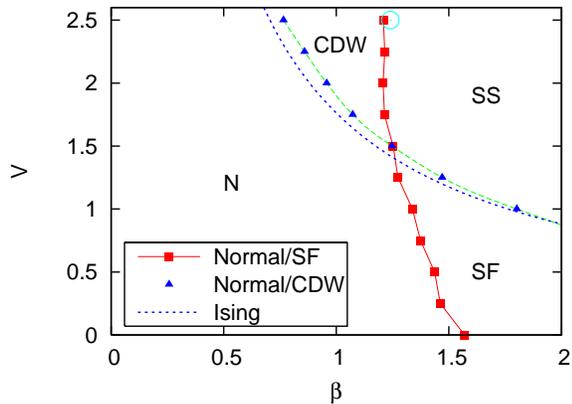,angle=  0,width=8cm}}
\caption{\label{Phases}
The central result of our paper:
Phase diagram of the extended Feynman-Kikuchi 
model in the $V-\beta$ plane at half-filling. 
Squares denote simulation results for the superfluid transition, 
and triangles the CDW transition.
The CDW boundary is in good correspondence with the 
Ising limit $\beta_{c}^{\,{\rm cdw}} =
4/(2.269 \, V)$, indicated by the dotted line.
The circle at $V=2.5$ is the prediction
$\beta_c^{{\, \rm sf}}(\rho)= 
\beta_c^{\, {\rm sf}}(\rho=1)/\rho$.
See text for details.
}
\end{figure}

Proceeding in analogy with the winding $\langle W^2\rangle$, we measure
$S(\pi,\pi)$ 
for different lattice sizes and
do finite size scaling to determine
$\beta_c^{\,{\rm cdw}}$ for the CDW transition. 
Representative plots, for $V=2.5$, are shown in
Figs.~\ref{DD} and \ref{DS}.
Putting together such sweeps for different interaction strengths $V$
yields the density order-disorder (CDW) phase boundary in the $\beta-V$
plane at half-filling shown in Fig.~\ref{Phases}.

In fact, a number of aspects of the phase diagram of Fig.~\ref{Phases}
can be inferred by a mapping to the Ising model.  We note that
the interaction energy $E_v$ of Eq.~6 maps precisely to that
of the Ising model (the well-known equivalence of lattice-gas and
Ising models)
with the definition of the spin $S^z_i = 2 (\rho({\bf r}_i) - \frac12)$.  
The alternating occupied and empty site pattern caused by the repulsive
particle-particle interaction ($V>0$) corresponds 
to an antiferromagnetic arrangement in spin language.
The quantity $V/4$ plays the role of the exchange constant $J$.
The critical temperature of the Ising model given
by the Onsager solution $T_c = 2.269 J$ then implies a density
ordering $\beta_c^{\,{\rm cdw}} = 4/(2.269 \, V) = 1.763/V$.
We expect this result to be accurate for $V >> \beta$ 
where $E_v$ dominates over any effects on the site densities
which might be caused by the exchange term.
This curve is shown as the dotted line in Fig.~\ref{Phases}.

We can also make a qualitative argument for how the phase boundary might
bend away from this Ising limit as the role of $\beta$ becomes larger.
For any given density arrangement,
the exchange of particles provides additional configurations
of the system associated with permutations of the particle indices.
Such configurations have lower energy when the occupied sites
are adjacent.  Thus we expect that the exchange term will
favor ``ferromagnetic'' spin configurations, in competition
with the ``antiferromagnetism'' driven by $E_v$.
This suggests a lowering of the CDW transition temperature.
Just such an increase in $\beta_c^{\, {\rm cdw}}$ is seen in the phase diagram.
% Indeed, the breakaway from the Ising limit begins as the
% superfluid region, where exchange effects are presumably
% large, is entered.  
One may well ask whether there is
an extreme limit where the attraction between particles
due to the greater ease of exchange becomes so dominant that
phase separation occurs (ferromagnetic clusters in spin language).
We will address this possibililty later.

The corresponding effect of $V$ on the superfluid phase 
transition is less easy
to describe rigorously, in part because we do not begin from a known
limit like the Onsager solution in the density transition case.
On the one hand, increasing $V$ drives the particles apart,
making local exchange more expensive, suggesting that
$\beta_c^{\,{\rm sf}}$ might increase.  
On the other hand, the superfluid transition
is not caused by local exchange but instead by global winding, and 
by separating the particles, $V$ might help provide regularly
spaced ``stepping stones,'' aiding global winding and 
decreasing $\beta_c^{\,{\rm sf}}$.
While these qualitative arguments provide different conclusions,
numerically, the answer
is clear from Fig.~\ref{Phases}: turning on $V$
decreases $\beta_c^{\,{\rm sf}}$.
The effect is not large, however.

In the limit of large $V$ and half-filling, a perfect CDW phase
forms, which corresponds to a completely filled
square lattice with a lattice constant
$\sqrt{2}$ larger than the original lattice (and rotated by 45 degrees).
The squared distances in the FK exchange energy will be scaled 
up by a factor of two, and hence
$\beta_c^{\,{\rm sf}}$ will be twice the value for the no-vacancy
FK model of the previous section.  Thus 
$\beta_c^{\,{\rm sf}} = 2\,(0.62 \pm 0.01)$.
This value is shown as the circle at $V=2.5$ in Fig.~\ref{Phases}.

In order to develop a simple understanding of 
$\beta_c^{\,{\rm sf}}$ for general $V$, we 
consider a small (four site, two particle) cluster 
and enumerate completely the
allowed configurations.  
There are six possible density configurations, which separate
into two classes.  Four of them have the two particles adjacent,
while the other two have the two particles separated by vacancies:

\begin{eqnarray}
\begin{array}{l@{\;\;\;|\;} l@{\;\;\;|\;} l@{\;\;\;|\;} l@{\;\;\;|\;}}
 {\rm Configuration} & {\rm K \; energy} & {\rm P \; energy } &  {\rm Weight }  \\
 \hline
\bullet - \bullet - \circ - \circ & 1/2 + 9/2 & V & 4 \, e^{-\left(\frac{5}{\beta} + V \beta \right) } \\
 \hline
\bullet - \circ -\bullet  - \circ & 4/2 + 4/2 & 0 & 2 \, e^{-\left(\frac{4}{\beta} \right) } \\
\end{array}
\nonumber
\end{eqnarray}

\noindent
We have included the degeneracy factors in the weight.
For each density configuration, there are two permutations.
Since we are interested in the superfluid
transition, we will restrict ourselves to the case where the
two particles do exchange, which is reflected in the nonzero value of
the kinetic energy in the table above.
 
The expectation value of the Kinetic energy is
\begin{eqnarray}
 \langle K \rangle &=& \frac{ 2 \cdot 5 e^{-\left(\frac{5}{\beta} + V \beta \right) } + 4 e^{-\left(\frac{4}{\beta} \right)} }
   { 2  e^{-\left(\frac{5}{\beta} + V \beta \right) } +  e^{-\left(\frac{4}{\beta} \right)} } 
\nonumber
\\
 &=& 4 + \frac{2} { 2 + e^{\left(\frac{1}{\beta} + V \beta \right) } } 
\end{eqnarray}
Note that $\langle K\rangle = 4$ is kinetic energy at $V\rightarrow\infty$.
The superfluid transition occurs when 
$K \sim \beta$.
If we set $\langle K\rangle =\beta_c^{\,{\rm sf}}$ and use
$ \langle K(V\rightarrow\infty) \rangle = 4 =
 \beta_c^{\,{\rm sf}}(V\rightarrow\infty) $ we find that the
shift in the superfluid transition is given by,
\begin{eqnarray}
\frac{\Delta \beta_c}{\beta_c(V\rightarrow\infty)} = \frac{1}{2} \cdot \frac{1} { 2 + e^{\left(\frac{1}{\beta} + V \beta \right) } } 
\end{eqnarray}
This result 
is qualitatively correct at 
$V=0$, predicting a small positive shift in 
$\beta_c^{\,{\rm sf}}$ relative to $V \rightarrow \infty)$.

\subsection {Doped system  $\langle \rho \rangle \;\ne \; \frac{1}{2} $}

\begin{figure}[t]
\centerline{\epsfig{figure=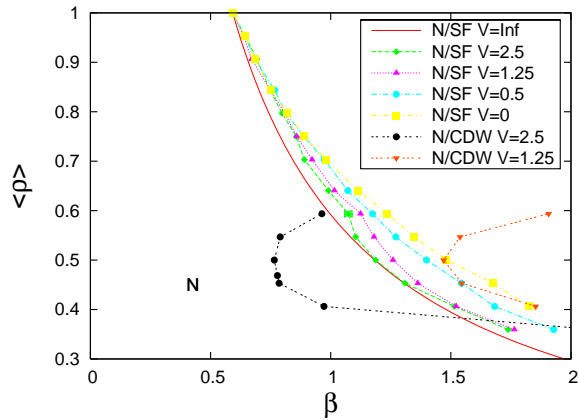,angle=  0,width=8cm}}
\caption{\label{Phases2}
Phase diagram in the $\rho-\beta$ plane for the extended FK
model for different values of $V$. 
Triangles:   the numerically obtained cdw transitions.
Squares:   the numerically obtained superfluid transitions.
The superfluid phase boundary is reasonably well approximated by
$\beta_c^{\,{\rm sf}}
\propto 1/\rho$ (solid curve).  See text.}
\end{figure}

In this section we consider general filling $\rho$.
Specifically, in Fig.~\ref{Phases2} we exhibit the
 phase diagram in the $\beta-\rho$ plane for two fixed values
of the interaction, $V=1.25$ and $V=2.50$.
Fig.~\ref{Phases2} was obtained using the same analysis as in
the earlier sections:  Evaluation and scaling of the
winding and structure factor as a function of $\beta$
for different $V$ and $\rho$.
Several features are immediately apparent from the phase diagram:
Charge ordering is, as expected, favored close to half-filling,
with the highest transition temperature at $\rho=1/2$.
Interestingly, the shape and size of the density ordered region
around half-filling is in rough agreement with the boundaries
obtained for checkerboard solid order
in the extended boson-Hubbard model \cite{batrouni00,hebert02} where
similar superfluid and charge ordered phases are present in 
an explicitly quantum model.

In the preceding section we argued 
that $\beta_c^{\, {\rm sf}}$
for half-filling and $V=\infty$ should
be a factor of two larger than for the original, no vacancy
FK model, and showed this was borne out numerically.
One might expect that a similar result
would be true for general fillings and that
$\beta_c^{\, {\rm sf}}$
would be increased by a factor
of $1/ \rho$, since this factor reflects the increase in the square
of the average interparticle spacing.
\begin{eqnarray}
\beta_c^{\,{\rm sf}}(\rho,V\rightarrow\infty)  =  
\frac{1}{\rho} \beta_c^{\,{\rm sf}}(\rho=1) 
\end{eqnarray}
However, on further consideration, it is not quite so.  Half-filling
and $V=\infty$ is a special case and in general
the particles are not dispersed
uniformly, that is, they no longer all have the same distance 
from their nearest neighbors.
Nevertheless, 
this relation provides a reasonable
guide to the density dependence of the
superfluid transition, and 
is shown on phase diagram at Fig. \ref{Phases2} as the line 
``$N/SF\; V={\rm Inf}$".
%% REMOVED LONG DISCUSSION OF rho DEPENDENCE.  THE TEXT
%% IS PRESERVED AT THE BOTTOM OF THE FILE BELOW \end{document}
%% REMOVED EMPIRICAL DISCUSSION OF CDW TRANSITION.  THE TEXT/FIG
%% IS PRESERVED AT THE BOTTOM OF THE FILE BELOW \end{document}
In Fig.~\ref{Phases2} we also see that as
$\beta$ is increased at $\rho=1/2$ and fixed $V=1.25$, we go from 
normal to superfluid to supersolid, where the phases
are labeled by the behavior of the 
two order parameters, $\langle W^2\rangle$
and $S(\pi,\pi)$.  Likewise at larger $V=2.50$ we go from
normal to CDW to supersolid.

While the order parameters provide
unambiguous identification of the phases, it is also interesting to
see if the thermodynamics can pick up the two successive transitions
in the form of separate peaks in the specific heat.
To study $C(T)$,
we proceed similarly to the derivation for the 
partition function with only the kinetic energy term.
Now our partition function has both potential and kinetic energy. 
\begin{eqnarray}
Z &=& 
 \summ_{n} e^{-P \beta - \frac{K}{\beta}} 
\nonumber\\
E &=& -\frac{\partial \ln Z}{\partial \beta} 
 = 
\frac{\summ_{n}\left(P-\frac{K}{\beta^2}\right) e^{-P\beta -\frac{K}{\beta}}}{Z_0}
\nonumber\\
C &=& \frac{\partial E}{\partial T} =\frac{\partial E}{\partial (1/\beta)} 
\nonumber\\
&=& 
- \left(\langle P\rangle \beta - \frac{\langle K\rangle}{\beta} \right)^2 
\nonumber\\
 &+& \left\langle \left(P \beta - 
\frac{K}{\beta} \right)^2 \right\rangle -\frac{\langle K \rangle }{\beta} 
\end{eqnarray}

\begin{figure}[t]
\centerline{\epsfig{figure=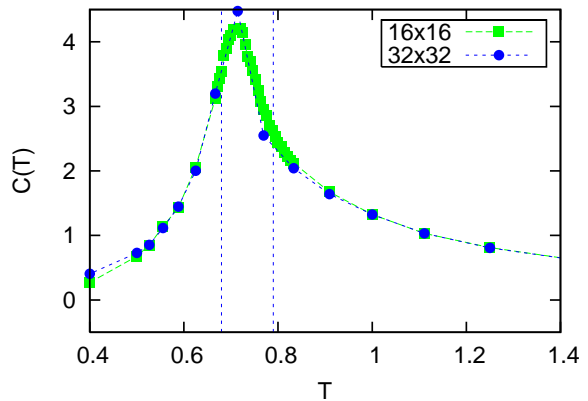,angle=  0,width=8cm}}
\caption{\label{CXV1.25}
Specific heat for half-filling, $V=1.25$.
From Fig.~\ref{Phases2},
the $SF$ transition is at $T=0.79$, and the $CDW$ transition at $T=0.68$.
These values are shown as vertical dashed lines.
The specific heat cannot resolve the two peaks.}
\end{figure}

Fig.~\ref{CXV1.25} shows the specific heat as a function of
temperature for 
half-filling and interaction $V=1.25$.
According to the phase diagram Fig.~\ref{Phases2},
the two transition points are
$N/SF$ at $\beta=1.26$ ($T=0.79$) and $N/CDW$ at $\beta=1.47$ ($T=0.68$).
As can be seen, we cannot resolve separate peaks in $C(T)$ 
associated with these transitions.
It is likely that the critical temperatures are too close
and that the finite size rounding blurs the two peaks into
a single maximum.

%% REMOVED CLOSE UP FIGURE

\begin{figure}[t]
\centerline{\epsfig{figure=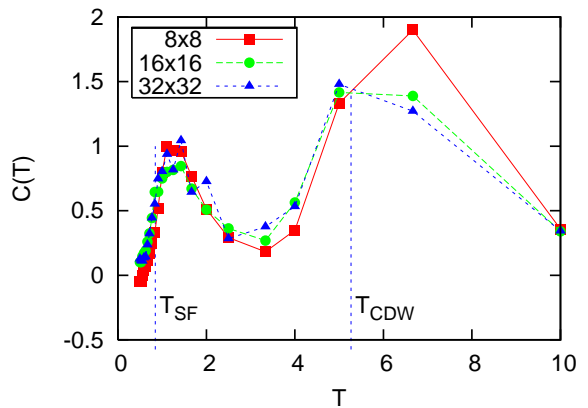,angle=  0,width=8cm}}
\caption{\label{CXV10}
Specific heat for half-filling and $V=10$.
Two peaks corresponding to SF and CDW transitions are
clearly seen.  
The critical values given by the winding and structure factor
are shown as vertical dashed lines.
The Ising mapping would give $T_{CDW} = 5.67$.
}
\end{figure}

We can however exhibit separate SF and CDW peaks in the specific heat if
we push the transitions apart sufficiently.
For example, at $V=10$ 
the CDW transition occurs at a much higher temperature
than the SF transition.  Indeed, in Fig.~\ref{CXV10}
we can now observe separate signatures of
the two transitions in $C(T)$.  The maxima occur
close to the transition points given by the order parameters.

Our final results concern the possibility of phase separation.
One might argue that in a model with vacancies, especially
at low or vanishing $V$, the particles will clump together
in order to facilitate exchange.  Indeed, phase separation has been
observed in a related model:  the bose Hubbard Hamiltonian
with ring exchange, precisely due to this mechanism \cite{rousseau04}. % *** ??

%% REMOVED NEW PHASE SEPARATION ORDER PARAMETER.  TEXT
%% IS PRESERVED AT THE BOTTOM OF THE FILE BELOW \end{document}

Phase separation is signalled by a peak in the density
structure factor at small momenta ${\bf q}$ (as opposed
to the CDW ordering vector at the largest ${\bf q}=(\pi,\pi)$).
Crudely speaking there are real space 
density fluctuations at long
wavelengths, corresponding to a lattice with one side half occupied
and the other half empty.  These translate into a peak in
$S({\bf q})$ at small ${\bf q}$.
Note that in a canonical ensemble simulation such as is performed 
here we cannot set ${\bf q}=(0,0)$ since that value
of the structure factor is just a constant set by the filling.
Indeed with our definition of the density correlations in terms of
fluctuations about the average density per site, Eq.~14, 
$S(0,0) = 0$.

% In Fig.~xx we show $S(q_x,q_x)$ as a function of $q_x$
% for different values of $\beta$ and $V=0$.  We see that
% as $\beta$ increases... **What do we see?**

\begin{figure}[t]
\centerline{\epsfig{figure=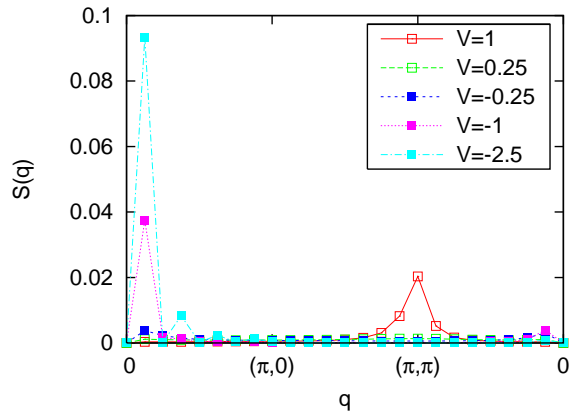,angle=  0,width=8cm}}
\caption{\label{PS-}
$S(\vec q)$ for various $V$, $L=16$, $\beta=1.5$.
$S(2\pi/L, 0)$ should signal phase separation
if it is $\sim \; 0.1$. It is however not close to that number for $V>0$.
For comparison, plots for $V<0$ are given, and they indicate phase separation.
}
\end{figure}

For a perfectly phase separated state with all particles on the
right-most half of the lattice
we find $S(2\pi/L, 0)\approx 0.11$ for $L=16$.
In Fig. \ref{PS-} we see that %$S(\vec q)$ 
$S(2\pi/L, 0)$ is $\sim \; 1/100$ of that figure for $V>0$.
We conclude there is no phase separation in this model.
The absence of a signal for phase separation is in contrast to the 
behavior in the related bose-Hubbard model with ring exchange 
\cite{rousseau04,rousseau05}
There the same quantity, the average of the structure factors at the three 
lowest momenta shows a sharp rise with increased exchange, as one enters
the superfluid phase.  

%% REMOVED SECTION ON COMPUTATIONAL RESTRICTIONS.  THE TEXT
%% IS PRESERVED AT THE BOTTOM OF THE FILE BELOW \end{document}

\section*{Conclusions}

In this paper we have presented Monte 
Carlo simulations of the phase diagram of an extension of the
$d=2$ Feynman-Kikuchi model which includes vacancies.
We found phases which have density and superfluid order, and where
these two types of order coexist.  Unlike the boson-Hubbard model
where supersolid order requires doping away from half-filling
%% \cite{oldlatticesupersims},
in the extended FK model $\rho_s$ is nonzero even in the
defect-free checkerboard solid.  The reason is that the boson-Hubbard
kinetic energy moves particles only between near-neighbor sites.
Bosons cannot exchange without passing through an energetically
unfavorable region.  But in the FK model, exchange at longer
range can occur without ever `passing through'
the rare configurations with near-neighbor sites
that are occupied.  One might expect that in the FK model
which is restricted to local exchange, the half-filled supersolid
might be eliminated.

A further problem of interest in the extended FK is to
consider ``quenched'' vacancies in which the locations of
the empty sites are frozen throughout the simulation.
Here again we might expect that when a restriction to local
exchange is enforced, there could be a destruction of the superfluid
transition as the percolation threshold is crossed.
In the model allowing exchanges of arbitrary distance, one
expects a more trivial increase in $\beta_c$, but that
the superfluid transition would likely persist.

A final avenue for exploration would be the inclusion of a
one-body vacancy potential which could be chosen to confine
the particles preferentially towards the center of the lattice.
Such simulations would connect with
recent experiments on cold atoms in magnetic and laser traps.

We acknowledge support from the National Science Foundation under award
NSF ITR 0313390, and useful input from T.Tremeloes.

\end{document}